\documentclass[12pt,preprint]{aastex}

\DeclareRobustCommand{\ion}[2]{%
\relax\ifmmode
\ifx\testbx\f
{\mathrm{#1\,\textsc{#2}}}\else
{\mathrm{#1\,\mathsc{#2}}}\fi
\else\textup{#1\,{\mdseries\textsc{#2}}}%
\fi}

\slugcomment{accepted to ApJ Letters October 16 2008} 

\shorttitle{Leading spiral arms in a LIRG}
\shortauthors{V\"ais\"anen et al.}

\begin{document}


\title{A pair of leading spiral arms in a luminous infrared galaxy? 
\altaffilmark{1}}  
\altaffiltext{1}{Based on observations made with the ESO telescopes 
at the
Paranal Observatory under programme 073.D-0406B and
observations made at the Anglo-Australian Telescope.}


\author{Petri V\"ais\"anen \altaffilmark{2}}
\altaffiltext{2}{South African Astronomical Observatory, P.O.Box 9,
       Observatory 7935, Cape Town, South Africa; petri@saao.ac.za}

\author{Stuart Ryder \altaffilmark{3}}
\altaffiltext{3}{Anglo-Australian Observatory, P.O.Box 296, 
Epping, NSW 1710, Australia; sdr@aao.gov.au}

\author{Seppo Mattila \altaffilmark{4}}
\author {Jari Kotilainen \altaffilmark{4}}

\altaffiltext{4}{Tuorla Observatory, Department of Physics and Astronomy, 
University of Turku, V\"ais\"al\"antie 20,  FI-21500 Piikki\"o, Finland;
seppo.mattila@utu.fi; jarkot@utu.fi}




\begin{abstract}
Leading spiral arms are a rare phenomenon.  We present here one of the 
very few convincing candidates of spiral arms opening counter-intuitively
in the same direction as the galaxy disk is rotating.  This 
detection in a luminous IR galaxy (LIRG) IRAS~18293-3134 is based on 
near infrared (NIR) adaptive optics imaging with the Very Large Telescope 
and long-slit NIR spectroscopy with the Anglo-Australian Telescope. 
We discuss the orientation of the galaxy based on imaging and derive rotation 
curves from both emission and absorption features in the spectrum. The galaxy 
is strongly star-forming and has a minor companion in a high-velocity
encounter.  
The fact that the arms of IRAS~18293-3134 
are not easily traceable from optical images
suggests that larger samples of high-quality NIR imaging of interacting
systems and LIRGs might uncover further cases of leading arms, placing
constraints on spiral arm theories and retrograde encounters, and especially
on the relationship between disk masses and dark matter halo masses.
\end{abstract}





\section{Introduction}

Though once a subject of active discussion, the sense of rotation
of spiral arms in galaxies has generally been considered as
resolved for half a century. In a thorough discussion
\citet{devauc} showed that all galaxies in
his sample had {\em trailing} arms, in agreement with the early position
of Slipher and Hubble, and in contrast to the {\em leading} arm camp 
of Lindblad.  

Afterwards the only systematic 
study of the subject with large samples of spirals is work done 
in the early 1980's 
by \citet[][]{pasha82} and \citet[][]{pasha85}.
The crucial question for them was not whether spirals are
trailing or leading, but rather whether {\em all} spirals are trailing, or
whether some leading arm cases could be found. 
In the end, they came up with a sample
of 4 leading arm galaxy candidates from a sample of close to 200.  However, one
of them (NGC~5395) was later shown not to be leading \citep{sharpkeel},
another (NGC~4490) is a highly dubious case regarding its orientation and 
existence of arms, and even the remaining two (NGC~3786 and NGC~5426) are
not clear-cut cases regarding their tilt, which determines the 
sense of rotation.
More recently, NGC~4622 has attracted attention 
\citep[][]{buta92,buta03,byrd08}.  
The authors conclude that regardless of its orientation, it has to
have leading arms since two pairs of detected arms have the opposite sense
with each other, in addition to single arms. Another example of a similar 
``counter-winding'' spiral structure in ESO~297-27 was recently published by
\citet{grouchy}. 

If leading arms are convincingly detected, it naturally places 
constraints on any theory of the origin and structure of spirals in galaxies. 
In the favoured density wave models 
\citep[see e.g.][]{binney} leading arms are allowed in principle: 
the density waves themselves can be trailing or leading, though trailing
waves are expected to be more robust \citep{toomre81}.
However, many theoretical studies suggest that retrograde encounters with 
companion galaxies should generate a robust single leading arm 
\citep{athanas, thomasson89, byrd93}. It is very interesting that indeed
every case of a suspected leading arm galaxy has a companion 
or a suspected companion. 
This then raises an interesting problem: 
since encounter directions are random,
it is not easy to understand why there are so very few observed candidates 
of leading arms (single or otherwise) in the multitude of observed 
interacting galaxies. \citet{thomasson89} find that if the disk mass is
similar to, or larger than the surrounding halo mass, the formation of a 
leading arm is suppressed in the interaction.  
The question to answer thus becomes whether 
i) the spiral galaxy dark matter (DM) halos really are so small; 
ii) whether leading arms are extremely short-lived (the modelling cited 
argues against this); 
or iii) are leading arms perhaps hidden from typical searches thus far? 

With the link to interactions and tidal disturbances and a way to probe the 
DM halo, leading arms are cousins of other intriguing and rare 
effects in disk galaxies, such as counter-rotating disks and (polar) rings 
\citep[][]{tremaine,vergani,brosch}. To understand spiral galaxy evolution, 
and to answer the questions above, more modelling and more and new kinds of 
observations are needed. 
In this Letter we present a strong candidate for the most clear-cut 
leading arm case to date, detected using a mode of observation
not attempted before in the case of leading arms: a galaxy with an obvious 
classical 2-arm spiral pattern as seen in adaptive optics (AO) imaging in 
the near-infrared (NIR).

\section{Observations and data reduction}

\begin{figure*}
\plotone{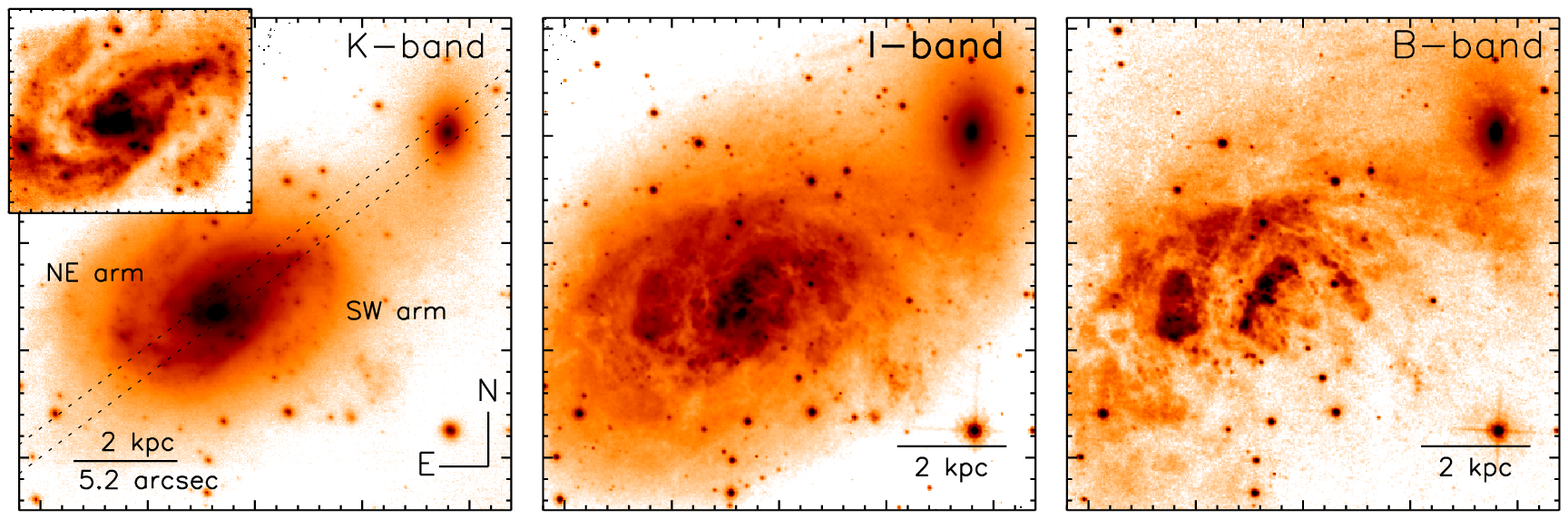}
\caption{The NACO image of IRAS 18293-3413 
         is at left, with the inset showing an un-sharp masked
         image to highlight the arms in better contrast. The 
        slit position observed with AAT/IRIS2 is indicated. 
         The HST/ACS $I$ and $B$-band images are at the centre 
         and right, respectively. 
  The Virgo infall-corrected distance to IRAS 18293-3413 is 
  79 Mpc making the angular scale 380 pc arcsec$^{-1}$.  
         Tick marks in the main panels are in one arcsec intervals,
         and all the brightness scales are logarithmic. \label{bigima}}
\end{figure*}

\object{IRAS 18293-3413}, a gas-rich spiral with an IR luminosity 
of $L_{IR} = 10^{11.7}{\rm L}_{\odot}$ \citep{sanders},
was observed with the NAOS-CONICA (NACO) AO instrument 
on the VLT UT4 as part of our program to find dust obscured 
core-collapse supernovae in 
the inner regions of LIRGs and to study star-formation and its triggering
in LIRGs \citep[][]{seppo07,petri}.
The $K$-band data set
used here was taken on 14 Sep 2004 with the S27 camera, giving 
a pixel size of 0.027 arcsec, and using the visual wave-front sensor.
These observations and data reduction are described in more detail in 
\citet[][]{seppo07}, where we presented the discovery of the highly obscured
core-collapse \citep[see][]{miguel07} supernova 2004ip detected within 
the nuclear regions of this galaxy. 
The final combined $K$-band image (Fig.~\ref{bigima}, left) 
has an on-source integration time of 1230 seconds. Two spiral arms are 
seen to open clock-wise around a single nucleus. 
We also extracted HST/ACS images from the archive (PI: Evans) which show 
the chaotic optical appearance of IRAS 18293-3413, though the arms can 
be traced in the $I$-band image aided by the NIR view.

Spectroscopic observations were obtained with the IRIS2 instrument at
the Anglo-Australian Telescope on the nights
of 27 and 28 Sep 2007, using a 1 arcsec wide slit, at a resolution 
R$\sim2400$.  
The position angle (PA = 128 deg) is close to the major axis of the galaxy 
and was such that the suspected companion galaxy to the NW also fell on the 
slit. 
Three sets of data were taken of the target, in the $J$, $H$, and $K$-bands,
in 300 sec nodded exposures.
Total integration time was 2700 sec in the $K$-band, and 3600 sec in the 
$J$ and $H$ bands.  An A0V-type standard star was observed in each band, 
as well as arc and flat lamps.

The AAT data were reduced with 
IRAF\footnote{IRAF is distributed by the National Optical Astronomy 
Observatories, which are operated by the Association of Universities for 
Research in Astronomy, Inc., under cooperative agreement with the National 
Science Foundation.} and IDL.
All science frames were flatfielded by a normalized smoothed ON-OFF flat, 
cosmic-ray corrected, and then subtracted pairwise from each other. Xe-arcs
were used to fit the wavelength solution and also correct for the 2D shape 
distortion of the frames.  A final background subtraction along columns was 
done to remove some sky-line residuals remaining after the pairwise 
subtraction. Frames were then shifted, co-added, and traced to produce
a clean, integrated, straightened 2D spectrum in each band.  
The standard star data 
were reduced in exactly the same way, except that the prominent hydrogen 
absorption features were fitted and removed.  The 1D extracted 
standard star spectrum was then divided into the target, and the result
multiplied by a smooth blackbody model of the star, thereby removing 
telluric features from the target spectrum while also performing a 
relative flux calibration.

\section{Analysis}

\subsection{Rotation curve}

To determine the sense of rotation of the galaxy we first analyzed the spectra
and derived rotation curves using five of the brightest emission lines
in the whole $JHK$ region. 
As an example Fig.~\ref{kspectrum} shows the extracted spectrum in
$K$-band over a central 4 arcsec wide aperture.
Single component Gaussians were fitted line by line 
in the 2D spectra
around the regions of interest, providing the wavelengths of the peaks.
The resulting heliocentric radial velocities as a function of
spatial position over the slit are plotted in 
Fig.~\ref{rot-emission}.  The radial velocities calculated
from all the lines are very consistent.
The rotation of IRAS~18293-3413 on the sky is 
such that the NW side facing the companion is approaching us.

We also examined the $^{12}$CO J=$2\rightarrow 0$ absorption feature
at 2.2935 $\mu$m in order to measure the velocity of the stellar
population in addition to the warm emitting gas producing the 
emission lines. For this purpose
we fitted stellar templates of giant and supergiant stars 
(Winge, Riffel \& Storchi-Bergmann, in prep.\
\footnote{http://www.gemini.edu/sciops/instruments/nir/spectemp/}) 
to the relevant $K$-band region (per 5 binned pixel rows) 
by redshifting the templates and convolving 
them with a range of internal velocity dispersions. The exact choice of the 
stellar template had no relevance for the resulting radial velocity, and 
only a small effect on the fitted velocity dispersion. 
The red overlay in Fig.~\ref{kspectrum} shows the
best-fit K7 III type template (HD63425B).
The resulting velocity curve is overplotted with
black circles over the emission line derived curves in 
Fig.~\ref{rot-emission}.  It is seen that the stellar rotation
follows closely that of the warm gas. 

The spectra also yield a redshift, for the first time, of the
smaller elliptical galaxy to the NW, 
based on the
$K$-band CO-feature and Fe II emission in $J$-band: we derived 
$5960 \pm 80$ km~s$^{-1}$, while the main galaxy systemic velocity 
is $5450 \pm 20$ km~s$^{-1}$.

\subsection{Orientation of the galaxy}
\label{orientation}

As thoroughly discussed in the literature 
\citep[e.g.][]{devauc,pasha85,sharpkeel},
a key issue in determining the sense of rotation
of a moderately inclined spiral galaxy 
is the difficulty in deciding its orientation in space 
(i.e.\ which side is nearer).
Two methods discussed in the literature are appropriate here: 
detection of dust lanes obscuring some of the brighter and smoother 
central regions of a galaxy often reveal the side of the disk nearer 
to the observer;
and essentially for the same reason the surface brightness profile along
the minor axis of the nearer side of the disk falls more abruptly than
the profile along the far side.

First, looking at the images in Fig.~\ref{bigima}, the significantly more 
drastic dust obscuration in the ACS $B$ and $I$ band images on the SW 
side of the system, and extending to S and SE, 
as well as the more pronounced dust filaments against the
brighter regions on the same SW side, give a visual impression of that 
being the near side of the galaxy.  More quantitatively, Fig.~\ref{profile}
shows the surface brightness profiles along the minor axis 
of the galaxy in all three images. Since the exact PA depends somewhat on the
isophote used, the plotted curves are averages from cuts ranging from 20 to
30 degrees. 
While the chaotic appearance and numerous star-forming regions make the 
profiles ragged 
it nevertheless is clear that the brightness on average falls more rapidly 
on the SW side in all bands, suggesting this is the side nearer to us.

\section{Discussion and conclusions}

\begin{figure}
\plotone{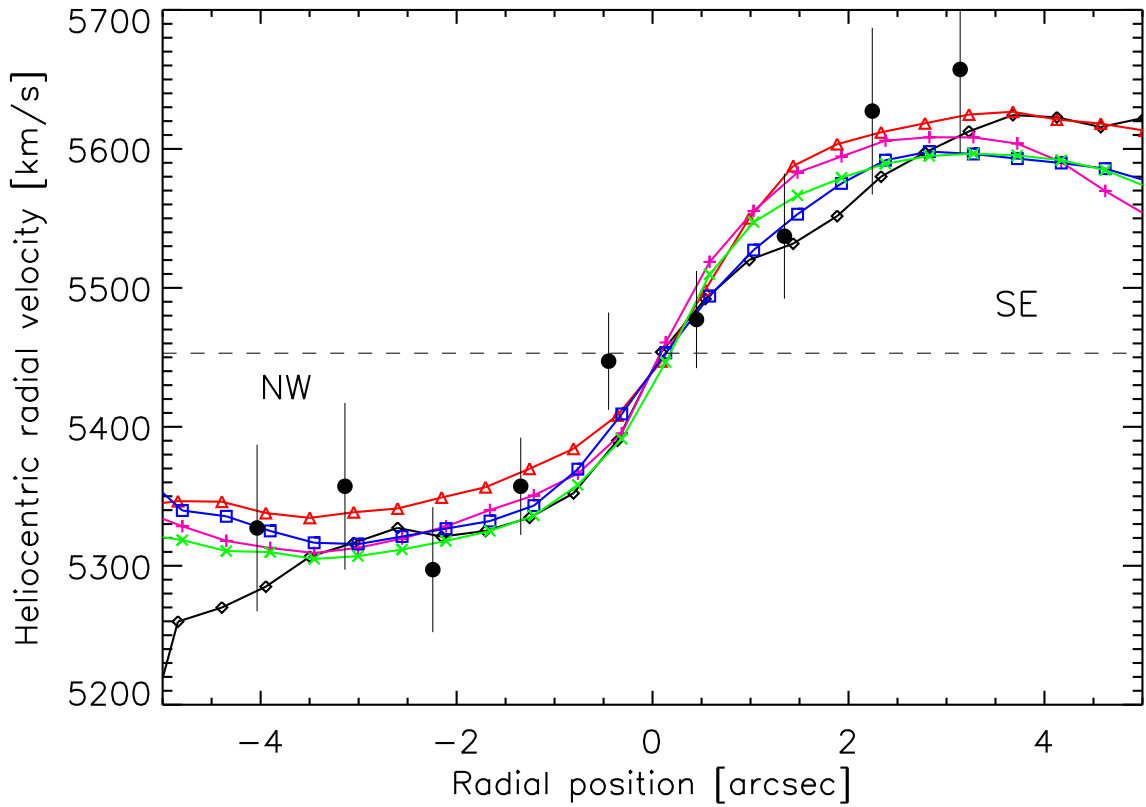}
\caption{The line-of-sight velocity curve determined from
CO 2.2935$\mu$m band-head (solid circles), and various emission 
lines: $[$FeII$]$ 1.2567$\mu$m (black diamonds), Pa$\alpha$ (red triangles), 
HeI (magenta plus-signs), H$_{2}$(1-0)S(1) (blue squares), 
and Br$\gamma$ (green crosses).
Error bars are not shown for the latter as they are on the order of
the scatter between the individual curves.
\label{rot-emission}}
\end{figure}

Figure~\ref{rot-emission} illustrates 
that the side of the galaxy closest to the companion
galaxy is the approaching side.  If we now accept, based on imaging and surface
brightness profiles, that the SW side of IRAS~18293-3413 is  
the one nearer to us, we have to conclude that the galaxy is turning
clockwise in the image, i.e.\ in the same direction that the spiral
arms open.  It is a leading arm spiral.

\begin{figure}
\plotone{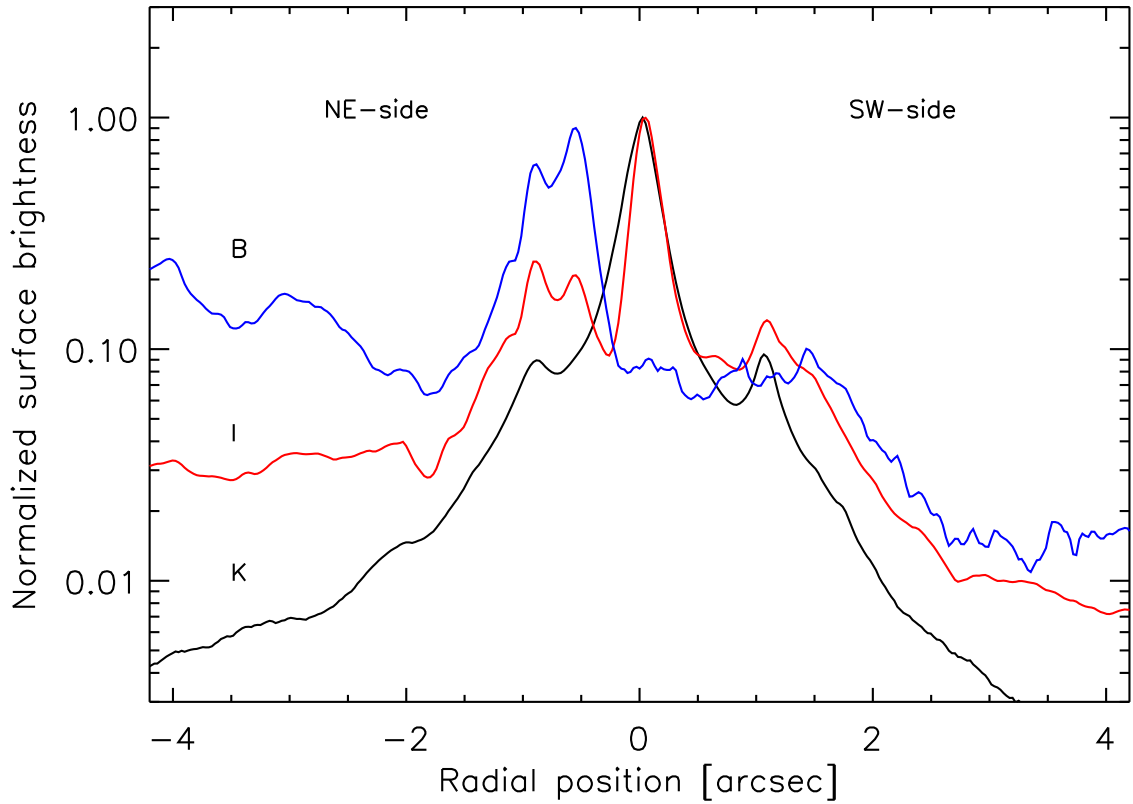}
\caption{Surface brightness profiles along the minor axis of IRAS 18293-3413 
         are shown in all three bands. The y-scales are normalized to maximum
         brightness in each band, while the x-axis is tied to the location
         of the $K$-band nucleus.  The brightness falls more rapidly in the
         SW-side of the galaxy. \label{profile}}
\end{figure}

Is there any ambiguity in the result leading to the classification of
a leading arm spiral? Since spectroscopy can 
be interpreted in only one way, the tilt of the galaxy opens the
only possibility of misinterpretation. 
It could be that while the SW side of the disk with the obscuring dust 
is the near one, the {\em stellar} 
disk and arm structure as seen in the NIR has 
the opposite tilt.  Though possible in principle, this is extremely 
unlikely: it would require that the rotational velocities and angles of the
de-coupled stellar and gas disks are 
{\em exactly} fine-tuned to produce the identical line-of-sight rotation 
curves measured (Fig.~\ref{rot-emission}). De-coupling in itself is not 
far-fetched at all as evidenced by counter-rotating disks. 
On the other hand, and more seriously, perhaps the significant amount of 
extinction on the SW side of the galaxy (Fig.~\ref{profile}) is not related
to the main galaxy and its orientation at all, but is rather foreground
material e.g.\ stripped off of a companion galaxy.  
Given the chaotic nature of LIRGs and the assumed interaction here, 
this is quite plausible -- indeed it may not be trivial to apply `normal' 
tilt-determining criteria to LIRGs.
If this were the case, the tilt of the galaxy would be
more difficult to judge, and would have to be based essentially only on 
NIR imaging. 
However, since our slit in the spectral observations does overlap the dust 
structure on the SE and NW sides of the galaxy we would expect some signal 
of de-coupled and/or complex velocity fields 
\citep[compare e.g.][]{petri}. We see no such structures. Moreover, the visual
impression of the $I$-band image does suggest the intertwining of the dust
structures and the stellar light clearly associated with the spiral arms
especially on the E and SE sides of the 
galaxy.  Nevertheless, with the present data we cannot resolve this ambiguity, 
and rather stress that we have detected {\em a very 
strong candidate} for a leading arm spiral.  Future integral field unit 
spectroscopy, or long-slit observations at other orientations could put the
issue to rest.

Based on our knowledge of peculiarities in spiral galaxies, the origin
of the leading arms is likely
due to an encounter with another galaxy.
The smaller elliptical (it is well fit by a de Vaucouleur profile)
companion is the obvious candidate, which makes
IRAS~18293-3413 a 'mixed' interacting/merging pair \cite[e.g.][]{phjoh08}, 
rather than a typical LIRG case of a pair of gas-rich spirals 
(unless the companion is the remaining bulge of a long-gone larger 
spiral). The radial 
velocity offset of $~500$ km s$^{-1}$ is surprisingly large, 
and typical only right at first 
{\em close} passages of merging galaxies
\citep{murphy,petri}.  However, given the disturbed morphology of the main
galaxy, an on-going LIRG-scale star formation burst, 
and that there are no other
2MASS galaxies seen within $7^{\prime}$ 
($\sim160$ kpc distance at $z=0.0185$), it is
very unlikely that the companion would be unrelated.  
Of course we do not know 
the transverse motion of the companion, 
but with the measured rotation of the main galaxy, the 
positive radial velocity offset of the companion is retrograde with 
respect to the rotation of the spiral disk.  This
fits well the general picture that it is retrograde encounters that might
trigger leading arms \citep[e.g.][]{thomasson89}. 

\begin{figure}
\plotone{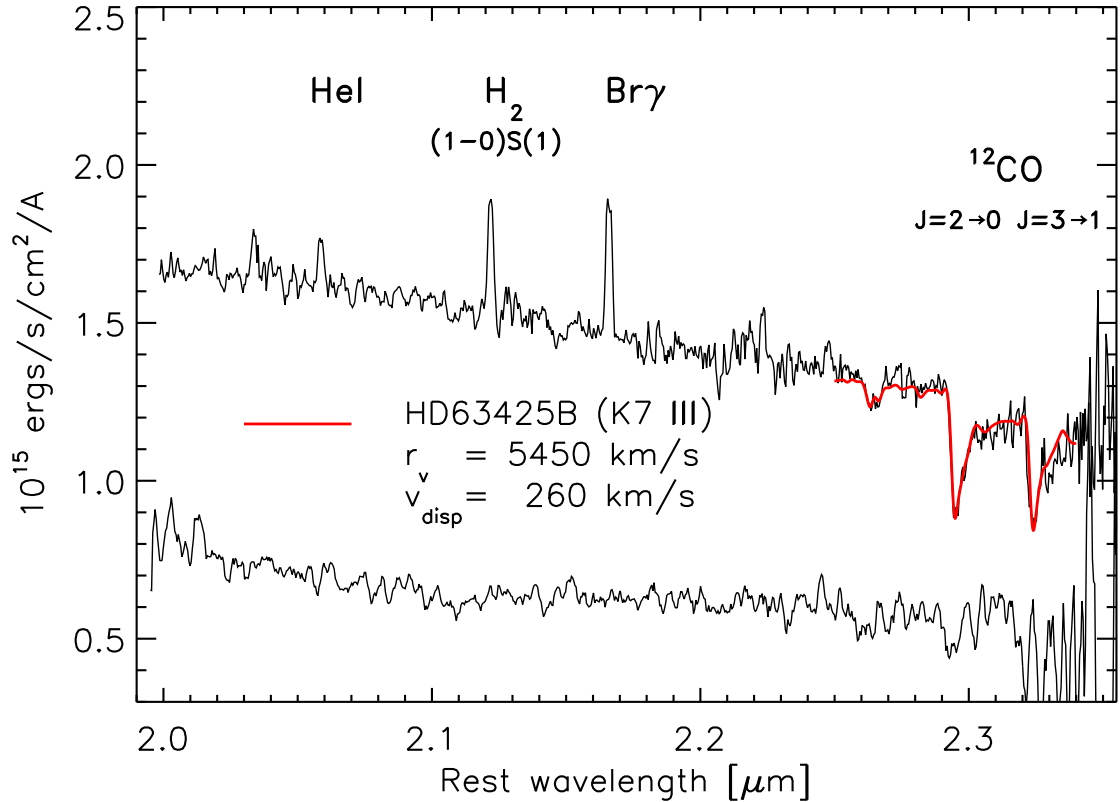}
\caption{The de-redshifted
$K$-band spectrum of IRAS 18293-3413 is shown at 
the top on a relative flux scale.
The main emission lines are labelled,
and a spectral template of a K7 III star is overplotted (thick red curve) 
in the region containing the CO band-heads. The bottom spectrum shows the 
companion galaxy, arbitrarily scaled and de-redshifted by 5960 km/s.
The LIRG spectrum is unsmoothed, while the companion spectrum is 
smoothed with a 4-pixel box-car. 
\label{kspectrum}}
\end{figure}

The leading arm simulations of \citet{byrd93} 
concerning NGC~4622 favour
a small companion of 1:100 mass-ratio 
passing close to a disk galaxy centre, while 
\citet{thomasson89} show how a more distant and massive 
retrograde companion produces a leading arm. 
Our system clearly resembles more the small perturber case.
We can estimate the galaxy masses
from their NIR light and velocity dispersion. 
We measure $M_{K} \approx -21.4$ for the companion, while the main galaxy 
has $M_{K} \approx -25.0$. Assuming both have the same
mass-to-light ratio, their mass-ratio would thus be close to 1:30.
The companion does however have a fairly significant mass of  
$\sim7\times10^{9} \ {\rm M}_{\odot}$ based on the 
$K$-band mass-to-light ratio of \citet{thron}. 
The CO-bands are detected in the companion, though 
with low signal-to-noise ratio (Fig.~\ref{kspectrum}). 
By directly fitting a Gaussian to the feature, and correcting for instrumental 
effects, we arrive at $\sigma \approx 90$ km/s.
Using an effective radius of $r_e = 0.62$ kpc determined by surface
brightness profile fitting with GALFIT \citep{peng},
we calculate a dynamical mass estimate \citep[e.g.][]{petri} of
$\sim7\times10^{9} \ {\rm M}_{\odot}$,  
while an equivalent mass estimate for the main galaxy results 
in $\sim1.1\times10^{11}\ {\rm M}_{\odot}$ (adopting 50 deg inclination, 
observed $v_{rot}=140$ km/s, $r_e=1.0$kpc, $\sigma=260$).  The ratio is
1:16, though given the uncertainties of the $\sigma$
determination of the companion the result is consistent with the 1:30 ratio
from NIR light.

Perhaps the most far-reaching aspect of leading arms is their connection to
cosmology via DM halos --  these halos are 
generally thought to dominate the dynamics of galaxies, but their nature and 
structure are still largely unknown \citep[e.g.][]{ostriker,olling,sofue}.
\citet[][]{thomasson89} found that leading arms form only
when the surrounding DM halos dominate the disk mass. The non-existence of
leading arms could thus be seen as support for maximum-disk and small 
DM halo models, while more evidence, such as our result, for leading arms 
would support massive DM halos.  

There are not many candidate 
leading arm spirals in the literature. Of the handful, the case of
IRAS~18293-3413 is unique: the interpretation is significantly less 
ambiguous than any of the \citet{pasha85} candidates, and ours is less
complex compared to the better known case of NGC~4622 and the recent 
ESO~297-27. The latter two are ``counter-winding spirals'' with sets 
of arms opening in opposing directions, while our case appears to be a 
classical 2-arm spiral (in the NIR). 
Notably, this is the type of a leading spiral that has not been 
theoretically studied in the literature 
to the degree that single leading arms have.
In particular, whether a 1:30 mass-ratio companion in a high-velocity 
encounter described above actually can produce a pair of leading arms, 
and what the arms would imply about the DM halo properties, 
obviously needs to be followed up by modelling.

Typical to many LIRGs, but in contrast to the previous leading-arm candidates, 
IRAS~18293-3413 has a large dust and gas mass and is very strongly 
star-forming, and its appearance in the optical is chaotic.
We argue that this last aspect may in fact prove significant:
there appear to be no published systematic studies of spiral arm 
characterisation {\em and} rotation determinations {\em in the infrared}.
Once larger interacting IR-galaxy samples are followed-up
with high spatial resolution imaging in the NIR together with  
high-quality spectroscopy, we speculate that more leading arms 
may well be found. These cases must be searched for, 
and if more are identified, they should be studied and modelled 
dynamically. This will be crucial to fully understand the dynamics and 
evolution of disk galaxies.

\acknowledgments

We gratefully acknowledge useful discussions with Noah Brosch, 
Karl-Johann Donner, Peter Johansson and Mauri Valtonen. SM and JK acknowledge
financial support from the Academy of Finland (projects: 8120503 and 8121122).

{\it facilities:} \facility{VLT:Yepun}, \facility{AAT}

\clearpage

\begin{thebibliography}{100}

\bibitem[Athanassoula(1978)]{athanas} Athanassoula, E., 1978, A\&A, 69, 395
\bibitem[Binney \& Tremaine(1987)]{binney} Binney, J. \& Tremaine, S., 1987
{\em Galactic Dynamics}, Princeton University Press, 
Princeton NJ 
\bibitem[Brosch et al.(2007)]{brosch} Brosch, N., et al., 2007, 
\mnras, 382, 1809 
\bibitem[Buta, Crocker \& Byrd(1992)]{buta92} Buta, R., Crocker, D. \&
Byrd, G.G.,  1992, \aj, 103, 1526 
\bibitem[Buta, Byrd \& Freeman (2003)]{buta03} Buta, R.J., Byrd, G.G. \&
Freeman, T., 2003, \aj, 125, 634
\bibitem[Byrd, Freeman \& Howard(1993)]{byrd93} Byrd, G.G., Freeman, T. \& 
Howard, S., 1993, AJ, 105, 447
\bibitem[Byrd et al.(2008)]{byrd08} Byrd, G.G., Freeman, T., Howard, S. \& 
Buta, R.J., 2008, \aj, 135, 408
\bibitem[de~Vaucouleurs(1958)]{devauc} de Vaucouleurs, G., 1958, 
\apj, 127, 487
\bibitem[Grouchy et al.(2008)]{grouchy} Grouchy, R.D., Buta, R., Salo, H.,
Laurikainen, E., \& Speltincx, T., 2008, AJ, 136, 980
\bibitem[Johansson et al.(2008)]{phjoh08} Johansson, P.H., Naab, T. \& 
Burkert, A., 2008, ApJ submitted, arXiv:0802.0210
\bibitem[Mattila et al.(2007)]{seppo07}
Mattila, S., et al. 2007, \apj, 659, L9
\bibitem[Murphy et al.(2001)]{murphy} Murphy, Jr., T.W., Soifer, B.T., 
Matthews, K., Armus, L. \& Kiger, J.R., 2001, \aj, 121, 97 
\bibitem[Olling \& Merrifield(2000)]{olling} Olling, R.P. \& Merrifield, M.R.,
2000, MNRAS, 311, 361
\bibitem[Ostriker et al.(1974)]{ostriker} Ostriker, J.P., Peebles, P.J.E. \&
Yahil, A., 1974, ApJ, 193, L1 
\bibitem[Pasha \& Smirnov(1982)]{pasha82} Pasha, I.I. \& Smirnov, M.A.,
1982, Astrophys. and Space Sci., 86, 215
\bibitem[Pasha(1985)]{pasha85} Pasha, I.I., 1985, Sov.Astron.Lett., 11, 1 
\bibitem[Peng et al.(2002)]{peng} Peng, C.Y., Ho L.C., Impey C.D., \& 
Rix H.-W., 2002, AJ, 124, 266
\bibitem[P\'erez-Torres et al.(2007)]{miguel07} P\'erez-Torres, M.A., et al.
2007, \apj, 671, L12
\bibitem[Sanders et al.(2003)]{sanders} Sanders, D.B., Mazzarella, J.M., 
Kim, D.-C., Surace, J.A., \& Soifer, B.T. 2003, \aj, 126, 1607
\bibitem[Sharp \& Keel(1985)]{sharpkeel} Sharp, N.A. \& Keel, W.C., 1985,
\aj, 90, 469 
\bibitem[Sofue \& Rubin(2001)]{sofue} Sofue, Y. \& Rubin, V., 2001, ARAA, 
39, 137
\bibitem[Thomasson et al.(1989)]{thomasson89} Thomasson, M., Donner, K.J.,
Sundelius, B., Byrd, G.G., Huang, T.-Y., \& Valtonen, M., 1989,
A\&A, 211, 25
\bibitem[Thronson \& Greenhouse(1988)]{thron} Thronson Jr, H.A. \& 
Greenhouse M.A., 1988, \apj, 327, 671
\bibitem[Toomre(1981)]{toomre81} Toomre, A., 1981, in Structure and
Evolution of Normal Galaxies, ed. S.M.Fall \& D.Lynden-Bell, Cambridge
University Press, Cambridge, p. 111
\bibitem[Tremaine \& Yu(2000)]{tremaine} Tremaine, S. \& Yu, Q. 2000, \mnras, 
319, 1
\bibitem[Vergani et al.(2007)]{vergani} Vergani, D., Pizzella, A., 
Corsini, E.M., van Driel, W., Buson, L.M., Dettmar, R.-J., Bertola, F.,
2007, A\&A, 463, 883
\bibitem[V\"ais\"anen et al.(2008)]{petri} V\"ais\"anen, P., et al. 2008,
\mnras, 384, 886

\end{thebibliography}
\end{document}